\documentclass[10pt, journal]{IEEEtran}

\usepackage{float}
\usepackage{caption}
\usepackage{color}

\usepackage{cite}
\usepackage{amsmath,amssymb,amsfonts}
\usepackage{amsthm}
\usepackage{graphicx}
\usepackage{textcomp}
\usepackage{xcolor}
\usepackage{svg}
\usepackage{subfigure}
\usepackage{algorithm}
\usepackage{algorithmic}

\newtheorem{theorem}{Theorem}

\newtheorem{lemma}{Lemma}

\newtheorem{corollary}{Corollary}

\allowdisplaybreaks
\makeatletter
\def\ScaleIfNeeded{%
\ifdim\Gin@nat@width>\linewidth \linewidth \else \Gin@nat@width
\fi } \makeatother
    
\graphicspath{{fig/}}

\begin{document}

\title{Multiple-Antenna Aided Aeronautical Communications in Air-Ground Integrated Networks: Channel Estimation, Reliable Transmission, and Multiple Access}

\author{{Jingjing~Zhao, Yanbo~Zhu, Kaiquan~Cai, Zhen Gao, Zhu Han, and Lajos~Hanzo}}

\maketitle
\begin{abstract}
To provide seamless coverage during all flight phases, aeronautical communications systems (ACS) have to integrate space-based, air-based, as well as ground-based platforms to formulate aviation-oriented space-air-ground integrated networks (SAGINs). In continental areas, L-band aeronautical broadband communications (ABC) are gaining popularity for supporting air traffic management (ATM) modernization. However, L-band ABC faces the challenges of spectrum congestion and severe interference due to the legacy systems. To circumvent these, we propose a novel multiple-antenna aided L-band ABC paradigm to tackle the key issues of reliable and high-rate air-to-ground (A2G) transmissions. Specifically, we first introduce the development roadmap of the ABC. Furthermore, we discuss the peculiarities of the L-band ABC propagation environment and the distinctive challenges of the associated multiple-antenna techniques. To overcome these challenges, we propose an advanced multiple-antenna assisted L-band ABC paradigm from the perspective of channel estimation, reliable transmission, and multiple access. Finally, we shed light on the compelling research directions of the aviation component of SAGINs. 
\end{abstract}

\section{Introduction}
Aeronautical communications systems (ACS) constitute one of the key infrastructure elements of ensuring the safe and efficient operation of air transportation. According to the specific safety levels, aeronautical communications can be categorized into forecabin communications, represented by air traffic control (ATC) as well as airline operation control (AOC), and in-cabin communications, represented by air passenger communication (APC) \cite{Doc9718, Hanzo}. Naturally, forecabin communications must have top priority in aeronautical communication systems, given its significance in air traffic management (ATM) for ensuring efficient airspace traffic flows and flight safety. With the rapid development of the air transportation industry, forecabin communications tend to have ever-increasing requirements on the aeronautical communications capacity \cite{Hanzo}. However, the currently in-use narrowband forecabin communications relying on high frequency (HF), very high frequency (VHF) and aeronautical satellite communications (SATCOM), fail to cope with the rapid growth of data volume in the modern ATM systems. For example, the air-to-ground (A2G) data rate of the Aircraft Communication Addressing and Reporting System (ACARS) based on VHF is as low as $2.4$~kbps, while the data rate of the VHF data link mode 2 is a meagre $31.5$~kbps~\cite{Doc9718}. Therefore, there is an urgent need to develop high-performance aeronautical broadband communications (ABC).

\begin{figure}
    \centering
    \includegraphics[scale=0.31]{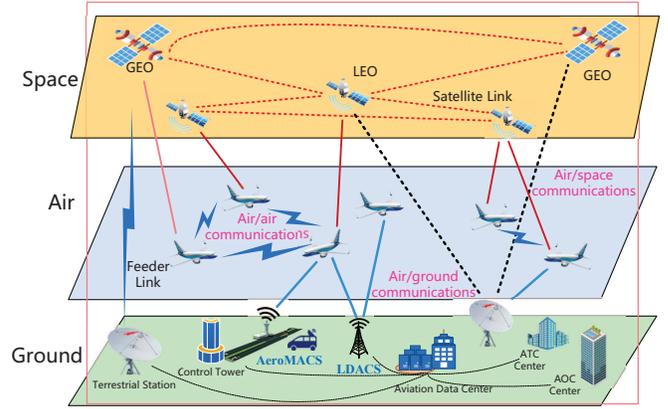}
    \caption{Aviation-oriented SAGIN structure.}
    \label{fig:SAGIN}
\end{figure}

\textcolor{black}{The International Civil Aviation Organization (ICAO) put forward the Aviation System Block Update (ASBU) plan in the fourth edition of the Global Air Navigation Plan (GANP) in 2013}, which pointed out that the new-generation ABC is mainly constituted by the Aeronautical Mobile Airport Communications System (AeroMACS), the L-band Digital Aeronautical Communications System (LDACS) and the next-generation maritime satellite communications system of Fig.~\ref{fig:SAGIN}. Specifically, AeroMACS has to support flight-data delivery near the airport, while LDACS is the main A2G data link for covering populated areas. Moreover, satellite components are utilized for oceanic and other remote areas.
The inherent integration of satellite, aerial and terrestrial systems can guarantee seamless connections for all flights.
The LDACS supports broadband A2G data transmission in the continental area, which was first proposed by EUROCONTROL and the Federal Aviation Administration (FAA) in 2009. The LDACS is based on the forth-generation (4G) mobile communications technologies and adopts orthogonal frequency division multiplexing (OFDM) to support data transmission in the Internet Protocol Suite (IPS)-based Aeronautical Telecommunications Network (ATN). The ICAO officially launched the standardization of LDACS in 2016, and approved the draft standard document in 2018, which recommended the specific implementation of LDACS~\cite{SARPS}. In 2019, ICAO released the LDACS white paper, in which the specific application scenarios and services of LDACS were provided. ICAO plans to officially release the standard document, as well as its verification reports and system manuals in 2022.

\begin{figure}
    \centering
    \includegraphics[scale=0.43]{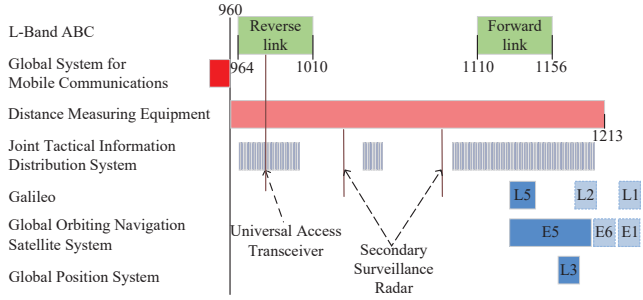}
    \caption{Spectrum occupancy status for L-band $960$ MHz-$1164$ MHz.}
    \label{fig:L-band-spectrum}
\end{figure}

Although the standardization of L-band ABC is being actively promoted, there are still many pending challenges to be solved. The ICAO stipulated that the deployment frequency band of LDACS is $960$~MHz-$1164$~MHz in its Doc 9718 standard document~\cite{Doc9718}. However, as shown in Fig.~\ref{fig:L-band-spectrum}, the spectral resources are scarce and fragmented in the frequency range assigned due to legacy systems such as the \textcolor{black}{Distance Measuring Equipment (DME)}, Joint Tactical Information Distribution System (JTIDS), etc. \textcolor{black}{Specifically, DME is the primary user of this frequency band, which relies on channels having the bandwidth of $0.5$~MHz and the spacing of $1.0$~MHz.} According to the ICAO recommendation~\cite{Doc9718}, the L-band ABC system will be embedded between the DME channels and occupy only $0.5$~MHz bandwidth. 
\textcolor{black}{Due to the out-of-band radiation, the DME system introduces non-negligible adjacent-channel interference. Although flight trials demonstrated by German Aerospace Center (DLR) have shown that LDACS performs well despite DME interference as a benefit of careful cell planning~\cite{DLR-flight-trials}, interference mitigation measures are still critical in scenarios associated with high DME activity.}
Therefore, the strong interference and scarce spectral resources are the two main challenges for the L-band ABC.

\textcolor{black}{With the fast development of public mobile communications, multiple-antenna techniques have become a mainstream trend as a benefit of their compelling merits, including high spectrum efficiency (SE), interference mitigation, power conservation, etc~~\cite{multiple-antenna, SAGIN-antenna}. As a consequence, it is also promising to apply multiple-antenna techniques in L-band ABC for addressing the problems of strong interference and scarce spectrum, as mentioned above.} However, due to the distinctive propagation environment of the L-band ABC, the application of multiple-antenna techniques possess specific challenges in terms of channel estimation, reliable transmission and multiple access, which motivates us to contribute this article. 

The rest of this article is organized as follows. We first identify the key features of the L-band ABC, and specify the technical challenges of applying multiple-antenna techniques. Then we propose a novel multiple-antenna aided L-band ABC paradigm from the perspective of channel estimation, reliable transmission, and multiple access. Finally, the open issues and research directions of the multiple-antenna assisted aviation-oriented SAGINs are discussed, and the article is concluded.
\section{Key Challenges for the Application of Multiple-Antenna Techniques in the L-Band ABC}
\label{sec:system-model}
In this section, we first specify the peculiarities of the L-band ABC propagation environment from the following four aspects.

\begin{figure}[!t]
\centering
\subfigure[]{\label{fig:f-domain}
\includegraphics[scale=0.5]{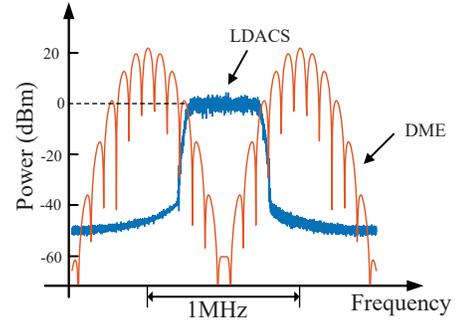}}
\subfigure[]{\label{fig:t-domain}
\includegraphics[scale=0.5]{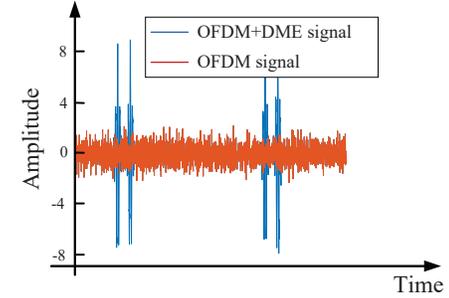}}
\caption{(a) The L-band ABC system is embedded between the DME channels; (b) DME interference is short and random in the TD.}\label{fig:LDACS-channel}
\end{figure}

\textbf{Strong interference:} In the frequency domain (FD), the L-band ABC system is embedded between the DME channels, as shown in Fig.~\ref{fig:f-domain}. Due to the out-of-band radiation, the spectrum of the L-band ABC system overlaps with that of the DME, hence imposing interference by the \textcolor{black}{DME} signals on the L-band ABC is inevitable. According to the research of the DLR~\cite{4702777}, under typical interference scenarios, the signal-to-interference power ratio (SIR) of the L-band ABC receiver may be as low as $-3.8$~dB. 
In the time domain (TD), the DME interference is short and random. On one hand, the duration of a DME pulse pair ($12$~\textmu s or $30$~\textmu s) is much shorter than that of the L-band OFDM symbol ($102.4$~\textmu s)~\cite{4702777}. On the other hand, the emission frequency of the DME pulse pairs from both the ground and airborne terminals  fluctuates around an average value. That is, the emission interval of DME pulse pairs is random, and the occurrence time of the pulse pairs follows the Poisson distribution, as shown in Fig.~\ref{fig:t-domain}. Explicitly, the DME interference cannot be readily quantified, it can only be estimated through its statistical characteristics.

\textbf{Sparse A2G multipath components (MPCs):} In contrast to terrestrial communications where the line-of-sight (LoS) path is often blocked, the LoS component of the en-route A2G channel dominates with high probability, except for low-elevation scenarios, where the LoS path is easily blocked by obstacles like mountains and hills. During the en-route phase, the A2G scattered components are typically not isotropically distributed. For example, in~\cite{channel-modeling}, the scatters were assumed to be uniformly distributed within a beamwidth of $\beta = 3.5^{\circ}$.

\textbf{Grave Doppler:} The average flying velocity of the aircraft during the en-route phase is around $1000$~km/h. Under the L-band frequency, the Doppler frequency offset can reach $1$~kHz. For the L-band OFDM signal with subcarrier spacing of $9.77$ kHz~\cite{SARPS}, the proportion of the Doppler frequency offset may exceed the acquisition capability of the synchronization module and cause severe inter-carrier interference (ICI). Note that when the aircraft passes a ground station (GS), the polarity of the Doppler changes. This change does not happen abruptly but rather the Doppler decreases gradually with the decreasing projected distance of the aircraft, which causes the maximum Doppler rate when the projected distance is $0$. This means that the synchronization unit has to be able to deal with this worst-case scenario of the Doppler rate. 
Moreover, the nonisotropic distribution of scatters leads to the fact that the Doppler probability density function is only a part of the classical 2-D isotropic Doppler density function. 

\textbf{Three-dimensional (3-D) dynamic topology:} In contrast to the two-dimensional (2-D) terrestrial mobile communications network, the aircraft are distributed at different altitudes, which leads to a 3-D aviation topology. The 3-D topology makes the network structure more complicated, and may severely affect the efficiency of spatial division multiple access (SDMA). Furthermore, given the influence of complex random factors, such as thunderstorms in the airspace, there may be a deviation between the actual flight track and the planned one, which brings uncertainties into the propagation environment. This makes the L-band ABC different from other communication scenarios having fixed terminal trajectories, such as high-speed railway communications. 

Given the aforementioned unique features of the L-band ABC, we summarize the key challenges for the application of multiple-antenna techniques from the perspective of channel modelling and estimation, reliable transmission, and multiple access, respectively, which are outlined in some detail in the following.

\textbf{Channel modelling and estimation:} During the en-route phase, the time-varying A2G channels experience strong DME interference, which makes the A2G channels significantly different from the channels of terrestrial communications systems. Therefore, it is vital to analyze the unique A2G channel features in the space/time/frequency domain, and build the channel model accordingly, so as to establish the theoretical basis of channel estimation. On the one hand, due to the ultra-high Doppler frequency offset, the pilot signals polluted by adjacent subcarriers will deteriorate the accuracy of channel estimation. Moreover, the burst DME signals on the adjacent channels impose unexpected interference on the pilot signals, which further improves the difficulties of channel estimation. On the other hand, the rapidly time-varying A2G channels have short coherence time, which requires frequent channel estimation. Meanwhile, the high dimensionality of the multiple antennas increases the complexity of channel estimation. Therefore, it is essential to realize efficient channel tracking with the aid of limited pilot overhead.

\textbf{Reliable transmission:} \textcolor{black}{Due to the existence of high-power DME signals in adjacent channels, the reliability of the L-band A2G transmission will be severely degraded. As a remedy, multiple antennas are capable of facilitating simultaneous signal enhancement and interference mitigation via sophisticated signal processing in the spatial domain.} The research of multiple-antenna aided interference mitigation in terrestrial mobile communications has indeed reached maturity. However, there are still open questions on how to utilize multiple-antenna techniques to improve the reliability of L-band ABC given the challenges listed below. Firstly, multiple-antenna enabled spatial interference filtering requires iterative optimization of weight vectors at the transmitter and receiver sides, respectively. This high-complexity process creates a bottleneck in the face of the limited airborne computing power and the violently time-varying A2G channel. Secondly, since the ratio of the Doppler frequency offset to the OFDM subcarrier spacing is far beyond the acquisition and tracking range of the state-of-the-art synchronization modules at the receiver, sophisticated Doppler compensation is necessary for reliability enhancement in the L-band ABC.

\textbf{Multiple access technique:} \textcolor{black}{Given the limited L-band spectral resources, spectral-efficient multiple access techniques are required.} \textcolor{black}{The multiple-antenna aided SDMA ideally requires uncorrelated channels for each element, otherwise severe inter-element interference is inflicted, owing to their correlation. However, on the one hand, sparse A2G MPCs result in a low-rank multi-aircraft channel matrix, and the unique 3-D topology further increases the probability of strong channel correlation among aircraft at different altitudes. On the other hand, the high-speed flight results in rapid fluctuation of the correlation coefficients of different aircraft channels. Therefore, how to support adaptive multiple access in the face of dynamically fluctuating low-rank A2G channels is a major problem to be solved. Additionally, given the random nature of flight tracks, robust real-time online resource allocation schemes have to be conceived.}


\section{Proposed Multiple-Antenna Aided L-Band ABC Paradigm}
To overcome the aforementioned limitations, this section proposes an advanced multiple-antenna aided L-band ABC paradigm, which provides reliable high-rate A2G data transmission in dynamically fluctuating environments in the face of strong interference and scarce spectral resources. \textcolor{black}{Given the compact form-factor limitation of airborne antenna deployments, we propose to use a small number of airborne antennas, but a large antenna array can be employed at the GS.} Since the spatial multiplexing gain is limited in case of sparse A2G MPCs, the popular hybrid combination of analog and digital beamforming can be adopted for the GS antenna array for reducing the hardware cost. 

\subsection{A2G Channel Estimation and Tracking}

\subsubsection{High-Resolution Channel Estimation Based on Multi-Domain Sparsity}
To realize high-resolution channel estimation under severe ICI and DME interference, we advocate the frame structure of Fig.~4(a). \textcolor{black}{Specifically, the channel estimation stage is split into two steps. In the first step, short OFDM symbols are adopted for frequency offset estimation, where the ICI can be ignored as a benefit of large subcarrier spacing. We use null-subcarriers to observe the DME interference in the FD. By leveraging the TD sparsity of DME interference, TD DME samples can be fully reconstructed despite having a limited number of null-subcarriers by sparse detection adopting popular compressed sensing (CS) algorithms~\cite{6742716}. With the DME interference subtracted from the received signals, the subcarrier frequency offset can be determined by exploiting the TD correlation of adjacent OFDM symbols. In the second step, the ICI can be first removed by compensating the estimated frequency offset, which is followed by the DME detection and subtraction, so as to obtain the ``decontaminated" pilot signals with both the DME and ICI removed.} 
	
\begin{figure}[!t]
\centering
\subfigure[]{\label{fig:frame-structure}
\includegraphics[scale=0.5]{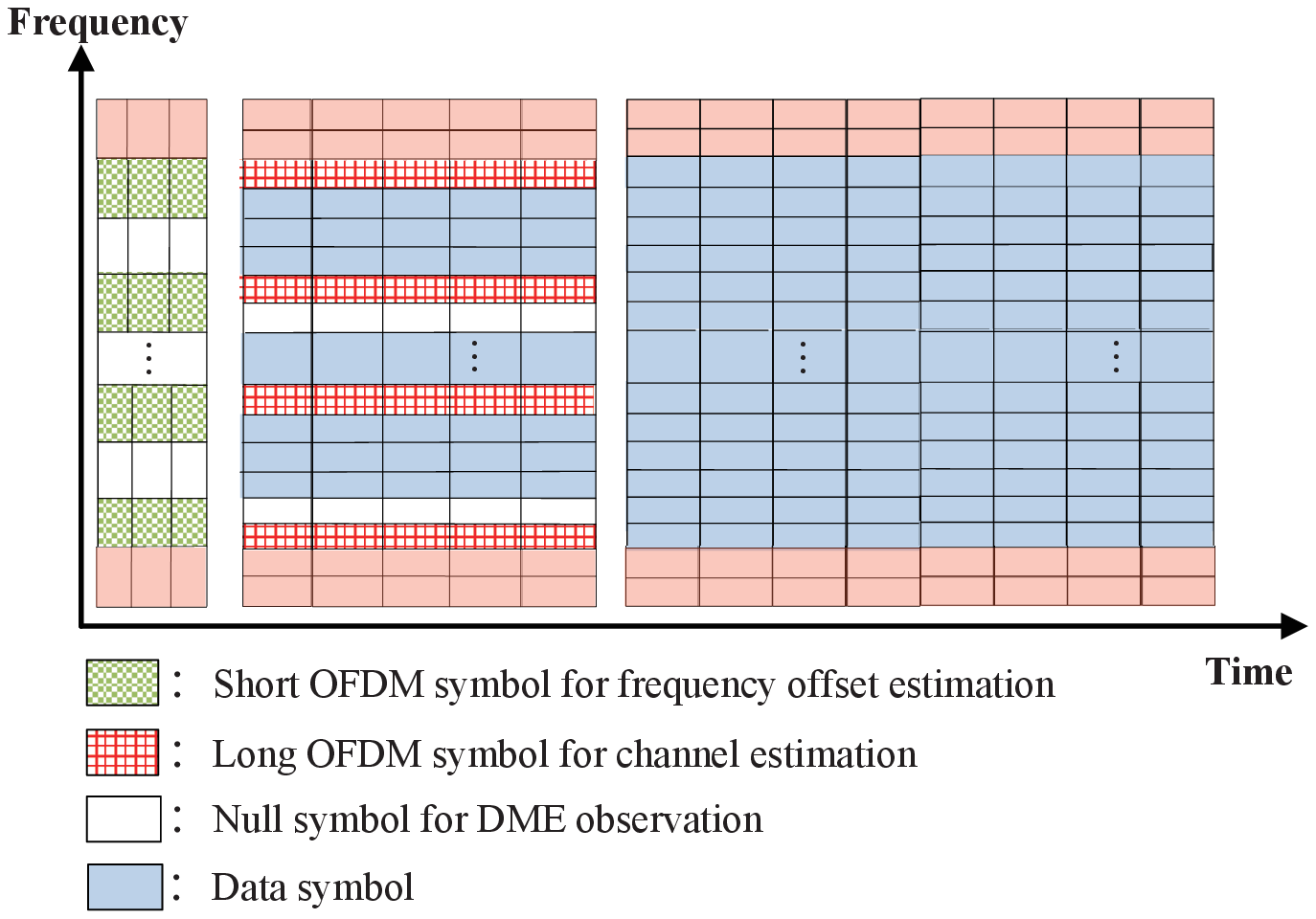}}
\subfigure[]{\label{fig:CE}
\includegraphics[scale=0.5]{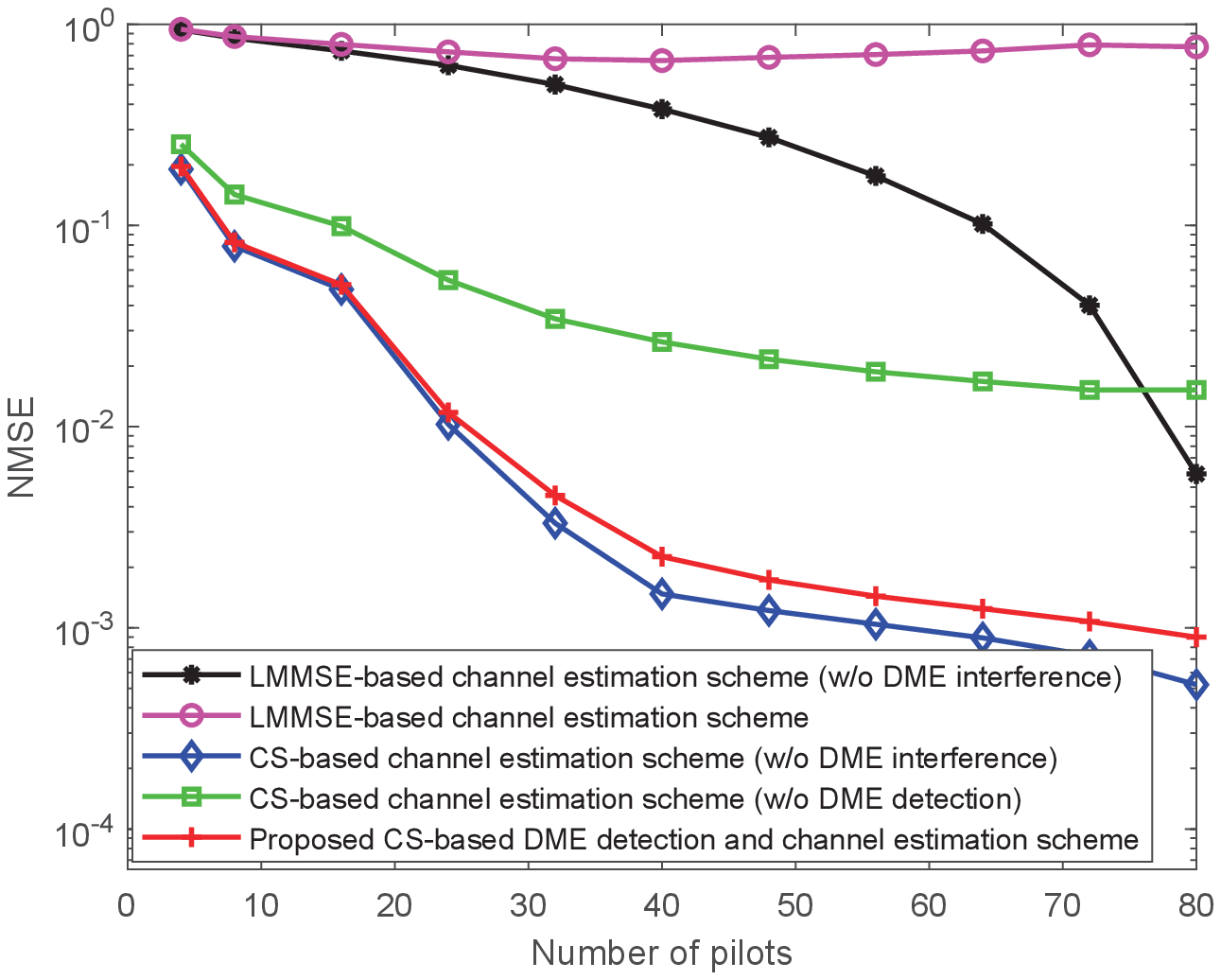}}
\caption{ (a) Frame structure conceived for channel estimation; (b) \textcolor{black}{NMSE performance comparison of different channel estimation schemes versus pilot overhead.}}\label{fig:chanenl-estimation}
\end{figure}

Given the common sparsity of the A2G channels over different subcarriers in the angular domain, the A2G channel estimation can be modeled as a generalized multiple measurement vector (GMMV) based CS problem~\cite{GMMV}. Then the GMMV based approximate message passing (GMMV-AMP) method may be harnessed for accurately reconstructing the A2G channels accurately at a low pilot overhead.
In Fig.~\ref{fig:CE}, we compare the proposed scheme with the conventional linear minimum mean square error (LMMSE)-based channel estimation scheme, where the normalized mean square error (NMSE) is chosen as the performance metric.
Evidently, the proposed scheme achieves much better performance than the LMMSE scheme, substantially mitigating the performance erosion caused by the DME interference.


\subsubsection{Flight Plan Assisted Channel Tracking}

\textcolor{black}{In aviation scenarios, following the regular flight track leads to predictable \textcolor{black}{spatial-domain} channel state information (CSI) based on the historical CSI~\cite{9598911}. This reduces the pilot overhead required for angular search at the GS. Moreover, given the flight plan information, the accuracy of the predicted spatial-domain CSI can be further improved by data fusion employing mathematical tools like Kalman filtering.} However, due to the random DME interference, it is difficult to directly track the high-dimensional channel. Therefore, we propose the following two-stage channel tracking scheme.

\begin{itemize}
\item \textit{Low-dimensional channel tracking}. 
\textcolor{black}{The low-dimensional channel $\hat{{h}}$ is first estimated with the DME removed by the airborne combiner.} Again, owing to the lack of scattering, the A2G channels exhibit sparsity in the delay domain. \textcolor{black}{Then, given the GMMV-AMP algorithm~\cite{GMMV}}, the low-dimensional channel can be accurately tracked using a relatively short training sequence.

\item \textit{High-dimensional channel tracking}.
\textcolor{black}{The high-dimensional channel $\hat{\pmb{H}}$ is tracked using pilot signals received at each airborne antenna. Given the predicted spatial-domain CSI, the complexity of the high-dimensional channel tracking is much reduced compared to the channel estimation stage. 
Then we convert $\hat{\pmb{H}}$ to the low-dimensional channel $\tilde{{h}} = \pmb{w}_A\hat{\pmb{H}}\pmb{w}_B$, where $\pmb{w}_A$ and $\pmb{w}_B$ represent the beamforming vector at the GS and the aircraft, respectively. By comparing $\hat{{h}}$ to $\tilde{{h}}$, we can confirm whether there exists DME interference during the channel tracking stage, thereby determining the reliability of the high-dimensional channel tracking results.}



\end{itemize}

\subsection{Interference Suppression and Doppler Compensation}\label{S:Beamforming}

\subsubsection{Dynamic Beam Management for Interference Suppression}
The multiple-antenna aided DME interference suppression has the following pair of challenges. Firstly, due to the randomness and sparsity of DME signals in the TD, it is hard to obtain accurate covariance estimates for the DME signals. Secondly, owing to the constant-modulus constraint of the analog beamforming at the GS side, the SINR-maximization problem becomes a non-convex non-deterministic polynomial (NP)-hard one. To tackle the above challenges, we propose a dynamic beam management strategy. Specifically, for the DME tracking, the initial statistical DME covariance can be inferred by the DME detection during the channel estimation stage. Since the real-time DME estimation requires high pilot overhead, we propose to utilize the powerful decision-feedback mechanism for DME tracking, where the DME covariance is updated based on the signals demodulated at the receiver, thereby improving the DME tracking efficiency. 

\begin{figure}
	\centering
	\includegraphics[scale=0.5]{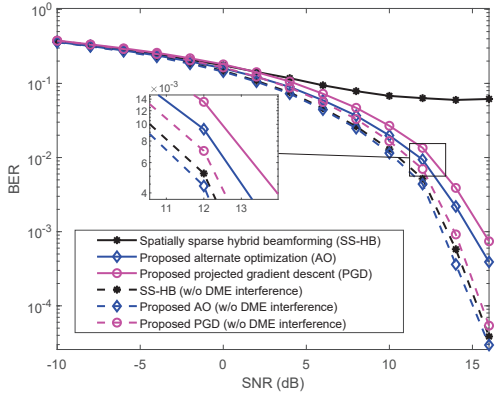}
	\caption{\textcolor{black}{BER performance comparison of different beamforming schemes versus SNR.}}
	\label{fig:AO}
\end{figure}

Given the statistical DME information, the subsequent transceiver beamforming can be optimized by the popular alternate optimization (AO) method, where the GS digital beamforming, GS analog beamforming, and the aircraft receive combining are decoupled into three subproblems. As the iterative approach is potentially time-consuming, which is less suitable for the rapidly fluctuating A2G channels, we propose to exploit the channel's correlation in the TD for low-complexity beam tracking. The projected gradient descent (PGD) algorithm is a promising technique of dynamically updating the beams by harvesting both the previous beamforming vectors and the current CSI. \textcolor{black}{In Fig.~\ref{fig:AO}, we compare the bit error rate (BER) of the proposed PGD scheme to that of the AO and the spatially sparse hybrid beamforming (SS-HB) schemes, when $4$ antennas are deployed at the aircraft. Observe that the BER of the proposed PGD scheme closely approaches that of the AO scheme, and outperforms that of the SS-HB scheme by substantially suppressing the DME interference.}

\subsubsection{Compensation of Ultra-High Doppler}
To overcome the ultra-high Doppler in high-speed flights, we propose a ``hierarchical compensation" scheme detailed as follows:
\begin{itemize}
    \item \textit{Rough compensation in the RF domain}. Based on the Doppler information gleaned during the channel estimation stage, the Doppler of the LoS path can be roughly compensated by the frequency shifters and phase-locked loop in the radio-frequency (RF) domain.
    \item \textit{Fine compensation in the baseband}. For the residual Doppler expansion of NLoS components that cannot be compensated in the RF domain, we propose to adopt the multi-scale wavelet transform in the baseband. Specifically, the different Doppler offset of MPCs in the FD corresponds to various scaling of signals in the TD. The matched filtering can be applied to the signals of each time scale, while the signals of other time scales are regarded as noise. After matched filtering, the signals are sampled in the TD according to their Doppler frequency offset to realize the frequency normalization. 
\end{itemize}

\subsection{Adaptive Broadband Multiple Access and Robust Resource Allocation}
Consider a typical scenario, where a GS transmits to $K$ aircraft. \textcolor{black}{The popular power-domain successive interference cancellation (SIC) technique is adopted for mitigating interference when SDMA gets ineffective due to highly-correlated A2G channels. Explicitly, we suggest to harness joint spatial-power domain multiple access to accommodate to the low-rank multi-aircraft channel matrix. To make the joint spatial-power domain multiple access adaptive to the rapidly fluctuating A2G channels, the K-means clustering algorithm - which has the merit of low complexity ($\propto K$) - is advocated for partitioning the aircraft having highly-correlated channels in the same cluster. Then, power-domain multiple access is applied in the same cluster, while SDMA is applied over different clusters.} 

\begin{figure}[!t]
\centering
\subfigure[]{\label{fig:uncertainty-set}
\includegraphics[width= 1.8in]{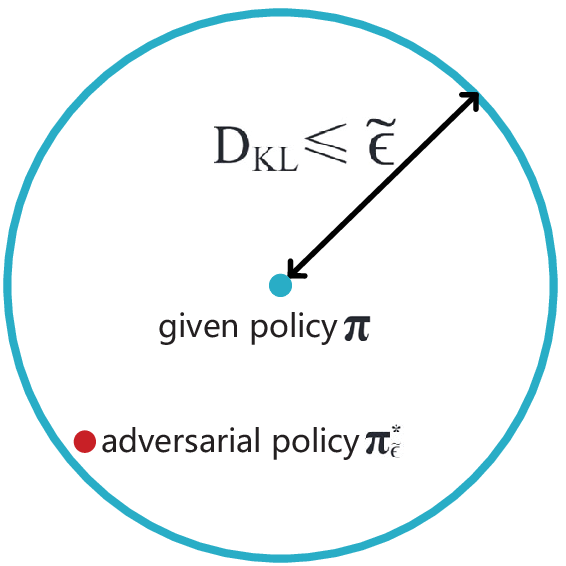}}
\subfigure[]{\label{fig:DRSAC-variance}
\includegraphics[scale=0.5]{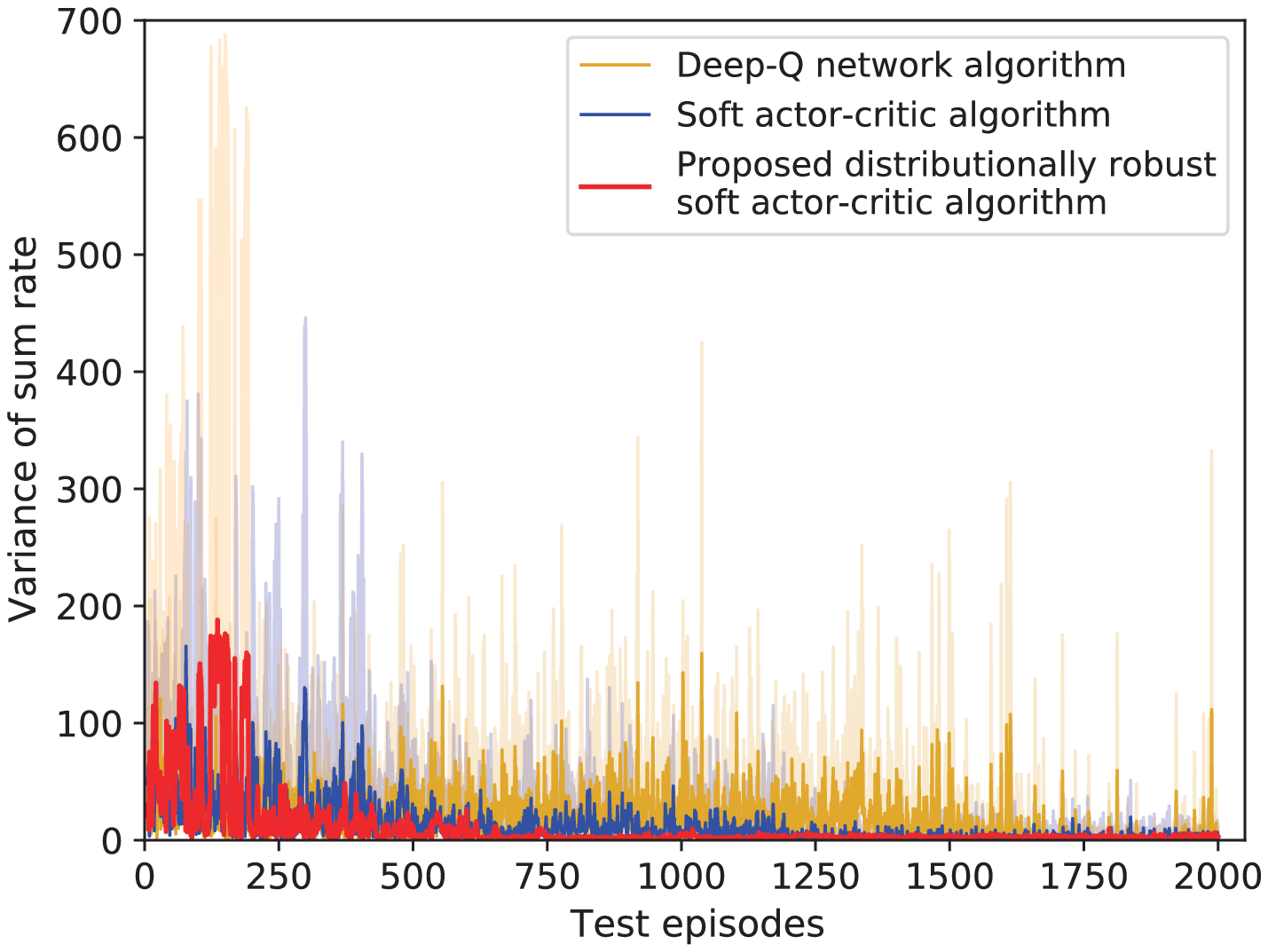}}
\caption{ (a) Uncertainty set of policies; (b) \textcolor{black}{Variance of sum rate achieved by DQN, SAC, DRSAC algorithms versus test episodes in the training phase~\cite{Jingjing-JSAC}.}}\label{fig:DRSAC}
\end{figure}

It is vitally important for the proposed adaptive broadband multiple access scheme to carefully optimize the real-time time/space/power-domain resource allocation for improving the A2G throughput. However, the real flight track may deviate from the planned one, which makes it impossible to fix the offline resource optimization for the entire flight-duration. For achieving high decision-making efficiency and robustness, we propose a ``distributionally-robust"\footnote{Inspired by the concept of distributionally robust optimization (DRO), an ambiguity set is introduced for characterizing the uncertain distribution of the policies, which is where the name of `DRRL' comes from. For more details, please refer to~\cite{drsac}.} reinforcement learning (DRRL) framework for solving the online resource allocation problem. In contrast to the conventional deep reinforcement learning (DRL), where the limited sampling of uncertain environments may lead to estimation errors, DRRL endogenously constructs the ambiguity set for capturing the uncertainties, and thus the worst-case performance is enhanced by avoiding any catastrophic outcome~\cite{Jingjing-JSAC}. Specifically, given a policy $\pi$ and error sequence $\Tilde{\epsilon} \in \mathbb{R}^{\mathcal{S}}$, the uncertainty set $\mathcal{U}_{\Tilde{\epsilon}}(\pi)$ can be constructed following the Kullback-Leibler (KL)-divergence~\cite{drsac}. The adversarial Bellman operator $\mathcal{T}^{\pi_{\Tilde{\epsilon}}^*}v(s)$ is defined as the minimum policy evaluation value following the policies within the uncertainty set~\cite{drsac}. As such, applying $\mathcal{T}^{\pi_{\Tilde{\epsilon}}^*}$ for the associated policy evaluation can provide the lower bound of state values, and thus prevent overly optimistic estimates. The uncertainty set construction is shown in Fig.~\ref{fig:uncertainty-set}. By leveraging the Lagrangian dual method, we can derive the expression of $\mathcal{T}^{\pi_{\Tilde{\epsilon}}^*}$ for robust policy evaluation~\cite{Jingjing-JSAC}. 

Referring to the above DRRL framework, we define the following Markov Decision Process (MDP):
\begin{itemize}
    \item \textbf{State}: The aircraft clustering result $\left\{\mathcal{G}_m\right\}$, the global CSI $\left\{\pmb{H}_{\text{FL},k}\right\}$;
    \item \textbf{Action}: Beamforming vector $\pmb{W}[x]$ at the GS, receiver combiner $\pmb{V}[x]$, power allocation $\pmb{p}[x]$, and time slot allocation vector $\pmb{\beta}_t$;
    \item \textbf{Reward}: The reward should be consistent with the objective, such as the sum rate.
\end{itemize}

Since the high-dimensional action space may result in unstable outcome, we propose to adopt the popular matching theory for solving the time slot allocation problem. Then we iteratively activate the matching algorithm relying on DRRL framework, until convergence is reached. Moreover, since the SIC decoding order within each cluster can be determined by the baseband equivalent channel, a near-optimal power allocation solution can be obtained by a low-complexity convex-optimization algorithm, which is embedded within the DRL framework. Therefore, the original action space $\left\{\pmb{W}[x],\pmb{V}[x],\pmb{p}[x],\pmb{\beta}_t\right\}$ is reduced to $\left\{\pmb{W}[x],\pmb{V}[x]\right\}$. For our multiple-antenna aided A2G scenario having uncertain flight trajectories, we have compared the performance of our distributionally-robust soft actor-critic (DRSAC) algorithm to the conventional deep-Q network (DQN) and soft actor-critic (SAC) algorithms in terms of the variance of the sum rate obtained over $5$ runs using different random seeds~\cite{Jingjing-JSAC}, as shown in Fig.~\ref{fig:DRSAC-variance}. As expected, the DRSAC algorithm substantially reduces the variance thanks to the robustness of the learning process.


\section{Open Issues and Research Directions}
The research of the multiple-antenna aided aviation-oriented SAGIN is still in its infancy, since many key technical issues are still open. In this section, we discuss several potential directions for future study.

\subsection{New Adaptive Modulation Techniques for High-Doppler Scenarios}
Again, OFDM is vulnerable to the ICI caused by the Doppler spread. Orthogonal time-frequency space (OTFS) modulation constitutes a promising design alternative to OFDM, which maps the information to the delay-Doppler domain~\cite{OTFS}. To apply OTFS in the aeronautical multiple-antenna system, low-complexity precoding and signal detection schemes have to be conceived for fully exploiting the spatial degree of freedom (DoF). Moreover, the design of resource allocation for OTFS-SDMA for multi-aircraft scenarios is also a challenging issue. 
\subsection{Spectrum Expansion}
In contrast to the over-crowded L-band, the millimeter-wave (mmWave) and TeraHertz (THz) bands have an abundance of under-utilized spectrum for supporting a high throughput. Although they exhibit high attenuation in terrestrial scenarios, at high altitudes they achieve higher propagation distances. 
Moreover, the extremely-short wavelength facilitates the deployment of numerous antennas on the aircraft body. Note that the airborne avionics system has access to multiple frequency bands, but its exploitation requires further study. Furthermore, the wide bandwidth and the high velocity result in complex delay, beam, and Doppler squint effects, causing inter-symbol interference (ISI) and undesired beam directions~\cite{THz}. By exploiting both the aircraft position and attitude, a prior information aided iterative channel estimation and beam alignment methods deserve further investigation. Another challenge is that in wideband scenarios, the Doppler of the lowest and highest subcarriers may be quite different, which makes its compensation a hard task.

\subsection{Seamless Handover}
The high mobility of aircraft leads to two types of handovers in aviation-oriented SAGINs, namely handovers among network segments, and intra-segment handovers, especially for trans-continental flights. More particularly, the aircraft can get connected via A2G links in the vicinity of airports or populated areas, but they have to switch to air-to-space (A2S) links in unpopulated oceanic or desert areas. As for handovers among access points within each network segment, these may represent switching between GSs in A2G communications or between satellites in A2S communications. Handover failures may inflict packet loss or even call drops, which is fatal for flight-safety. As the aircraft trajectories usually follow a limited set of predefined routes, efficient location prediction algorithms are essential for seamless handovers. Moreover, multi-objective Pareto-optimization is required for finding all optimal operating points of the non-dominated Pareto-front in terms of delay, power and BER minimization, as well as throughput maximization, for example.

\subsection{Routing Scheme}
The multi-layer SAGIN provides numerous routing options between the aircraft and terrestrial ATC/AOC centers. Specifically, trans-continental flights tend to require Pareto-optimal long-distance routing, all the way to the domestic AOC for ATM efficiency. Typically, there are numerous Pareto-optimal paths exhibiting extremely heterogeneous quality-of-service (QoS) requirements in terms of delay, power, BER and path-life-time, which underlines the need for sophisticated distributed learning-aided routing algorithms imposing low control overhead. 
The dynamic graph model and dynamic eigenvector theory constitute promising candidates for seamless network connectivity analysis and optimization. Moreover, due to the heterogeneity of the network segments, cooperative traffic-management and load-balancing also require frontier-research. The DRL technique is capable of tackling this challenge, using offline training and online deployment in the face of randomly fluctuating network traffic and link state at a low signaling overhead.




\subsection{Enhanced Applications}
The aforementioned techniques are particularly suitable for fundamental aviation applications, e.g., ATC, aircraft tracking etc. With the quickening pace of civil aviation evolution, the vision of formation flying and free flight is also gathering supporters. These emerging applications require extremely low latency, high reliability and robust communications for collision avoidance and accurate situational awareness. However, the long-distance A2G or A2S transmissions always suffer from large round-trip time. To facilitate the autonomous real-time reaction of aircraft, aeronautical ad hoc networks constitute a promising alternative to the conventional centralized technologies thanks to their self-organizing and self-discovering characteristics, especially in unpopulated areas. It has been shown in~\cite{Hanzo} that dozens of aircraft can be expected to be within the communication range of any aircraft at any time over the oceanic airspace. However, arranging multiple-antenna aided communication between aircraft is more challenging than A2G transmissions for the following two reasons. Firstly, the fast relative movement of aircraft leads to strict requirements concerning the timeliness of channel estimation, beamforming and Doppler compensation. Secondly, neighbor discovery is a challenge with narrow beam transmissions in highly-dynamic SAGINs. Therefore, the advanced beam tracking and hello message broadcasting using adjustable beamwidths are potential research topics.

\section{Conclusions}
SAGINs provide universal coverage for aircraft during all flight phases. Focusing on the future L-band A2G communications stipulated by ICAO, we have investigated the key challenges brought by the peculiarities of the L-band propagation environment and the ATM applications with an emphasis on multiple-antenna techniques. A sophisticated multiple-antenna aided L-band ABC paradigm has been proposed for channel estimation, reliable transmission, and multiple access, to support reliable high-rate transmissions in dynamic scenarios in the face of strong interference and scarce spectral resources. Our results have demonstrated the benefits of the proposed solutions. Finally, promising aviation-oriented SAGIN research directions have been discussed to help pave the way for designing the next-generation ABC.

\end{document}